\documentclass[reprint,superscriptaddress,
showpacs,preprintnumbers,nofootinbib,
 amsmath,amssymb,aps,floatfix
 ,prl
]{revtex4-1}
\usepackage{epsfig,dcolumn}
\usepackage{graphicx}
\usepackage{color}
\usepackage[utf8]{inputenc}
\usepackage{amsmath}
\usepackage{amsfonts}
\usepackage{amssymb}
\usepackage{xcolor}
\graphicspath{{figures/}}
\usepackage[linktocpage=true]{hyperref}
\usepackage{xspace}
\usepackage{newtxmath}
\usepackage{bm}

\newcommand{\be}{\begin{equation}}
\newcommand{\ee}{\end{equation}}

\newcommand{\sig}
{\ensuremath{\sigma/f_0(500)}\xspace}

\newcommand{\kap}
{\ensuremath{\kappa/K_0^*(700)}\xspace}

\newcommand{\pipikk}
{\ensuremath{\pi \pi \to K \bar K}\xspace}

\usepackage{bm} % scaled bold math symbols

\usepackage{mfirstuc} % to uppercase the name
\newcommand{\addReviewer}[2]{
  \expandafter\newcommand\csname #1\endcsname[1]{{\bf \color{#2} \capitalisewords{#1}:\,##1}}
  \expandafter\newcommand\csname #1cor\endcsname[2]{{\color{#2} \capitalisewords{#1}:\,\st{##1}{\bf ##2}}}
  \expandafter\newcommand\csname #1color\endcsname{#2}
}
\usepackage{soul,color}
\definecolor{chromeyellow}{rgb}{1.0, 0.65, 0.0}
\definecolor{DodgeBlue}{rgb}{0.118, 0.565,1.000}
\definecolor{asparagus}{rgb}{0.53, 0.66, 0.42}
\definecolor{cadmiumgreen}{rgb}{0.0, 0.42, 0.24}

\addReviewer{AR}{asparagus}
\addReviewer{JR}{red}

\newcommand{\wm}{Department of Physics, College of William and Mary, Williamsburg, VA 23187, USA}
\newcommand{\jlab}{Thomas  Jefferson  National  Accelerator  Facility, 12000 Jefferson Avenue, 
Newport  News,  VA  23606,  USA}
\newcommand{\ucm}{Departamento de F\'isica Te\'orica and IPARCOS, Universidad Complutense de Madrid, 28040 Madrid, Spain}

\begin{document}

\title{Determination of the lightest strange resonance $K_0^*(700)$ or $\kappa$, from a dispersive data analysis}

\author{J.R.~Pel\'aez}
\email{jrpelaez@fis.ucm.es}
\affiliation{\ucm}
\author{A.~Rodas} 
\email{arodas@wm.edu}
\affiliation{\jlab}
\affiliation{\wm}
%\date{\today}

\begin{abstract}
In this work we present a precise and model-independent dispersive determination from data of the existence and parameters of the lightest strange resonance \kap. We use both subtracted and unsubtracted partial-wave hyperbolic and fixed-$t$ dispersion relations as constraints on combined fits to $\pi K\rightarrow\pi K$ and $\pi\pi\rightarrow K\bar K$ data. We then use the hyperbolic equations for the analytic continuation of the isospin $I=1/2$ scalar partial wave to the complex plane, in order to determine the $\kappa/K_0^*(700)$ and $K^*(892)$ associated pole parameters and residues. 
\end{abstract}

\preprint{JLAB-THY-20-3166}
\maketitle
Despite the fact that Quantum Chromodynamics (QCD) was formulated almost half a century ago, some of its lowest lying states still ``Need confirmation'' according to the Review of Particle Properties (RPP) \cite{Tanabashi:2018oca}. This is the case of the lightest strange scalar resonance, traditionally known as $\kappa$, then $K^*_0(800)$, and  renamed $K_0^*(700)$ in 2018.

Light scalar mesons have been a subject of debate since the $\sigma$ meson (now $f_0(500)$) was proposed by Johnson and Teller in 1955 \cite{Johnson:1955zz}. Schwinger in 1957 \cite{Schwinger:1957em} 
incorporated it as a singlet in the isospin picture and pointed out that its strong coupling to two pions would make it extremely broad and difficult to find. A similar situation occurs when extending this picture to include strangeness and $SU(3)$ flavor symmetry. 
Actually, in 1977 Jaffe \cite{Jaffe:1976ig} 
proposed the existence of a scalar nonet below 1 GeV 
including a very broad \kap meson. 
Of this nonet, the $f_0(980)$ and $a_0(980)$ were easily identified in the 60's. However, the \sig and \kap have been very controversial for decades because they are so broad that their shape is not always clearly resonant or even perceptible.  Moreover, it was proposed \cite{Jaffe:1976ig,Jaffe:2007id} that these states might not be ``ordinary-hadrons'', due to their inverted mass hierarchy compared to usual quark-model $quark-antiquark$ nonets. In terms of QCD it has also been shown that their dependence on the number of colors is at odds with the ordinary one \cite{Pelaez:2003dy,Pelaez:2004vs,Pelaez:2015qba}. From the point of view of Regge Theory, a dispersive analysis of the \sig and \kap shows that they do not follow ordinary linear Regge trajectories \cite{Londergan:2013dza,Pelaez:2017sit}.

Definitely, both the \sig and \kap do not display prominent  Breit-Wigner peaks and their shape is often distorted by particular features of each reaction. Thus, it is convenient to refer to the resonance pole position, $\sqrt{s_{pole}}=M-i\Gamma/2$, which is process independent. Here $M$ and $\Gamma$ are the resonance pole mass and width. 
Note that poles of wide resonances lie deep into the complex plane and 
their determination requires rigorous analytic continuations.
This is the first problem of most \sig and \kap determinations: Simple models continued to the complex plane yield rather unstable results. In addition, many models assume a particular relation between the width and coupling, not necessarily correct for broad states, or impose a threshold behavior incompatible with chiral symmetry breaking. These are reasons why Breit-Wigner-like parameterizations---devised for narrow resonances--- are inadequate for resonances as wide as the \sig and \kap. 
In contrast, dispersion relations solve this first problem by providing the required rigorous analytic continuation. In practice, they are more stringent and powerful for two-body scattering.

The second problem is that  meson-meson scattering data are plagued with systematic uncertainties, since they are extracted indirectly from meson-nucleon to meson-meson-nucleon experiments. Inconsistencies appear both between different sets and even within a single set.  Thus, for very long, a rough  data description was enough for simple models to be considered acceptable, making model-based determinations of the \sig and \kap even more unreliable. Moreover, fairly good-looking data fits come out inconsistent with dispersion relations, as shown in \cite{Pelaez:2016tgi}. We will see here how they lead to very unstable \kap pole determinations.

Once again dispersion theory helps overcoming this second problem, by providing stringent constrains between different channels and energy regions. This explains the interest on dispersive
studies in the literature: for $\pi\pi$ \cite{Roy:1971tc,Ananthanarayan:2000ht,Colangelo:2001df,DescotesGenon:2001tn,Kaminski:2002pe,GarciaMartin:2011cn,Kaminski:2011vj,Moussallam:2011zg,Caprini:2011ky,Albaladejo:2018gif}, for $\pi N$ \cite{Steiner:1971ms,Ditsche:2012fv, Hoferichter:2015hva}, for $e^+ e^- \to \pi^+ \pi^-$ \cite{Colangelo:2018mtw}, for  $\gamma^{(*)} \gamma^{(*)} \to \pi \pi$ \cite{Hoferichter:2011wk,Moussallam:2013una, Danilkin:2018qfn, Hoferichter:2019nlq},
for $\pi K$ \cite{Johannesson:1976qp,Ananthanarayan:2000cp,Ananthanarayan:2001uy,Buettiker:2003pp,DescotesGenon:2006uk} and for $\pi\pi\rightarrow K\bar K$ \cite{Pelaez:2018qny} scattering. Actually, partial-wave dispersion relations implementing crossing correctly have been decisive in the 2012 major RPP revision of the $\sigma/f_0(500)$, changing its nominal mass from 600 to 500 MeV and decreasing its uncertainties by a factor of 5. In contrast, the \kap still ``Needs Confirmation'' in the Review of Particle Physics.

\begin{figure}
\includegraphics[width=0.48\textwidth]{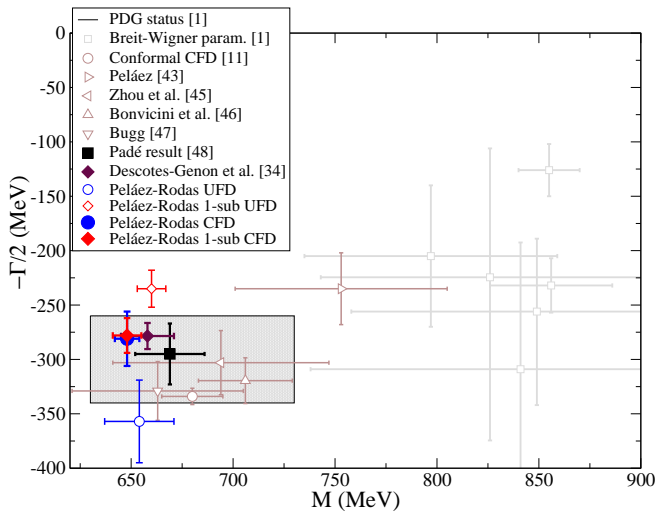}
 \caption{\rm \label{fig:poles} $K_0^*(700)$ pole positions.
 %Breit-Wigner parameterizations are taken from the RPP compilation \cite{PDG}, which also includes: Descotes-Genon et al. \cite{DescotesGenon:2006uk}, Bonvicini et al. \cite{Bonvicini:2008jw}, D.Bugg \cite{Bugg:2003kj}, J.R.Pel\'aez \cite{Pelaez:2004xp}, Zhou et al. \cite{Zhou:2006wm} and the ``Pad\'e Result'' \cite{Pelaez:2016klv}. The conformal CFD is our simple analytic extrapolation of our parameterization in \cite{Pelaez:2016tgi}.
 Selected from the Review of Particle Physics \cite{Tanabashi:2018oca,Pelaez:2016tgi,DescotesGenon:2006uk,Pelaez:2004xp, Zhou:2006wm,Bonvicini:2008jw,Bugg:2003kj, Pelaez:2016klv}
 We also show our results using Roy-Steiner equations, using as input
 our UFD or CFD parameterizations.
 Red and blue points use for $F^-$ a once-subtracted or an unsubtracted dispersion relation, respectively. 
 This illustrates how unstable pole determinations are when using simple fits to data.
 Only once Roy-Steiner Eqs. are imposed as a constraint (CFD), both pole determinations fall on top of each other. This final pole position is the main result of this work.}
\end{figure}

Note that a \kap pole is found as long as the isospin-1/2 scalar-wave data is reproduced and the model respects some basic analyticity and chiral symmetry properties \cite{vanBeveren:1986ea,Oller:1997ng,Oller:1998hw,Black:1998wt,Black:1998zc,Oller:1998zr,Close:2002zu,Pelaez:2004xp}. Furthermore, its pole must lie below 900 MeV \cite{Cherry:2000ut}. 
However, the pole position spread is very large when using models, as seen in Fig.\ref{fig:poles} which shows
the \kap poles listed in the RPP, together with a
shadowed rectangle standing for the RPP uncertainty estimate. Note that Breit-Wigner poles, displayed more transparently, have a very large spread and differ substantially from those having some analyticity and chiral symmetry properties built in.  

Some of those RPP poles used dispersive or complex analyticity techniques, although with approximations for the so-called ``unphysical'' cuts below threshold, which are the most difficult to calculate. 
In the $\pi K$ case, these are  a ``left'' cut, due to thresholds in the crossed channels, and a circular one due to partial-wave integration. Thus Fig.~\ref{fig:poles}  shows results from 
NLO Chiral Perturbation Theory (ChPT) unitarized with 
dispersion relations for the inverse partial-wave (Inverse Amplitude Method) \cite{Pelaez:2004xp} or for the partial-wave with a cut-off \cite{Zhou:2006wm}. In both cases the unphysical cuts are approximated with NLO ChPT.

The most sound determination 
of the \kap pole so far is the dispersive analysis by Descotes-Genon et al. \cite{DescotesGenon:2006uk}, also shown in Fig.~\ref{fig:poles}, which uses crossing to implement rigorously the left cut. The pole is obtained using as input
a numerical solution in the real axis of Roy-Steiner equations obtained from fixed-$t$ dispersion relations. Unfortunately these fixed-$t$ Roy-Steiner equations do not reach the \kap-pole region in the complex plane. 
Remarkably, in \cite{DescotesGenon:2006uk} it was shown that
the \kap region is accessible with partial-wave Hyperbolic Dispersion Relations. These were then used to obtain the pole, although starting from the solutions from the fixed-$t$ ones.
Note that this is a ``solving relations'' approach, since no data was used as input in the $\pi K$ elastic region around the nominal \kap mass, but only data from other channels and other energies
as boundary conditions to the integral equations.
In this sense, \cite{DescotesGenon:2006uk} provides a model-independent {\it prediction}. Despite this rigorous result, the \kap still ``Needs Confirmation'', and we were encouraged by RPP members to carry out an alternative dispersive analysis {\it using data}, as previously done by our group for the \sig \cite{GarciaMartin:2011jx}. The present work, which follows a ``constraining data'' approach instead of a ``solving relations'' approach, provides such an analysis.

It is worth noticing
that the \kap pole is consistent with $\pi K$ scattering lattice
calculations \cite{Dudek:2014qha,Brett:2018jqw}. For unphysical pion masses  of $\sim 400\,$MeV,
it appears as a ``virtual'' pole  below threshold  in the second Riemann sheet, consistently with expectations from unitarized NLO ChPT extrapolated to higher masses \cite{Nebreda:2010wv}. However, using pion masses between 200 and 400 MeV the pole extraction using simple models is again unstable \cite{Wilson:2019wfr}.
This makes the approach followed in the present work even more relevant, since in the future lattice will provide data at physical masses and energies around the \kap region that will require a ``constraining data'' technique for a model-independent description and its complex plane continuation.

Let us then describe our approach. In \cite{Pelaez:2016tgi} we first provided Unconstrained Fits to $\pi K$ Data (UFD) up to 2 GeV,
for partial waves $f^I_\ell(s)$ of definite isospin $I$ and angular momentum $\ell$,
 paying particular attention to the inclusion of systematic uncertainties. 
As usual, the total isospin amplitude $F^I(s,t,u)$, where $s,t,u$ are the Mandelstam variables, is the sum of the partial-wave series.
It was then shown that they did not satisfy well Forward Dispersion Relations (FDRs, $t=0$). However, we used these FDRs as constraints to obtain a set of Constrained Fits to Data (CFD), satisfying FDRs up to 1.6 GeV while still 
describing fairly data fairly well. Note that our ``conformal CFD'' parameterization of the low-energy isospin 1/2 scalar-wave already contains a \kap pole, shown in Fig.~\ref{fig:poles}. This is still a model-dependent extraction from a particular parameterization only valid up to $\sim1$ GeV. 

Later on \cite{Pelaez:2016klv}, we used sequences of Pad\'e approximants built from 
%derivatives of 
the CFD fit  to extract a new pole.
This ``Pad\'e result'' in Fig.~\ref{fig:poles} does not assume any relation between pole position and  residue, thus 
reducing dramatically the model dependence. The value came out consistent within uncertainties with the dispersive result in \cite{DescotesGenon:2006uk} and triggered the
recent change of name in the 2018 RPP edition from $K_0^*(800)$
to $K_0^*(700)$. However, it is not fully model independent, since the Pad\'e series is truncated and cuts are mimicked by poles.

Thus, we present here the \kap pole obtained from a full
analysis of data constrained to satisfy not only FDRs as in \cite{Pelaez:2016tgi}, but also both the $S$ and $P$
partial-wave (Roy-Steiner) dispersion relations. The latter are obtained either from fixed-$t$ or Hyperbolic Dispersion Relations (HDR),
along $(s-a)(u-a)=b$ hyperbolae, where $s,u$ are the usual Mandelstam variables. 
In \cite{DescotesGenon:2006uk} it was shown that the
convergence region of the latter in the $a=0$ case
reaches the \kap pole.

The price of using partial-wave dispersion relations is that they require input from the crossed channel $\pi\pi\rightarrow K\bar K$, whose partial waves  $g^I_\ell$ have the same two problems of being frequently described with models and the existence of two incompatible data sets (see \cite{Pelaez:2018qny} for details). In addition, there is an ``unphysical'' region between the $\pi\pi$ and $K\bar K$ thresholds, where data do not exist, but is needed for the calculations. Fortunately, Watson's Theorem implies that the phase there is the well-known $\pi\pi$ phase shift, which allows for a full reconstruction of the amplitude using the  Mushkelishvili-Omn\'es method.
Thus, in \cite{Pelaez:2018qny} we rederived the HDR partial-wave projections both for $\pi\pi\rightarrow K\bar K$ and $\pi K\rightarrow \pi K$, but choosing the center of the hyperbolas in the $s,t$ plane to maximize their applicability region.
Once again we found that simple fits to data do not satisfy well the dispersive representation, but we were able to provide constrained parameterizations of the two existing sets describing $S$-wave data up to almost 2 GeV, consistently with HDR up to 1.47 GeV within uncertainties. These are called CFD$_B$ and CFD$_C$ and are part of our input for the $\pi K$ HDR, although we have checked that using one or the other barely changes the \kap pole. Note that, contrary to previous calculations, we also provide uncertainties for $\pi\pi\rightarrow K\bar K$. Those for the $g^{1}_1$ wave are very relevant for the \kap pole, particularly in the unphysical region, where there is no data to compare with and the dispersive output leads to two different solutions when using one or no subtractions. Thus, we have also imposed in our CFD that the once and non-subtracted outputs should be consistent within uncertainties, which had not been done in previous calculations.

\begin{figure}
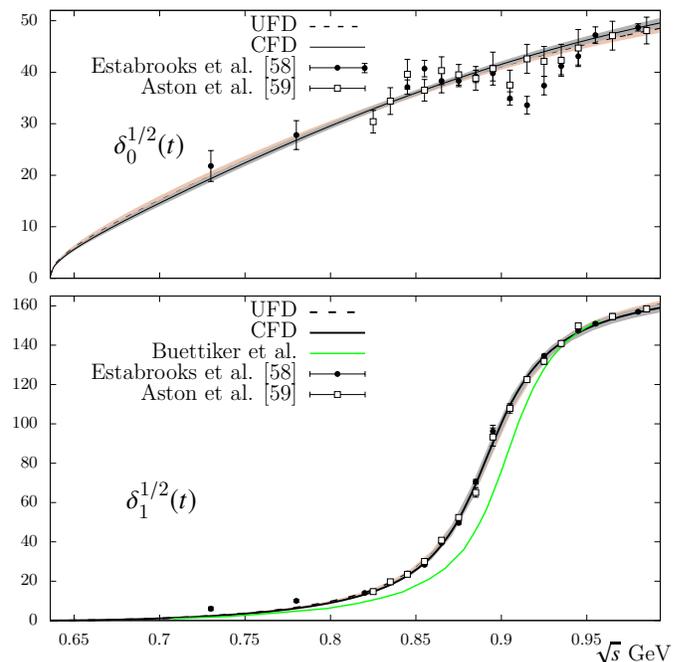

\resizebox{0.5\textwidth}{!}{\input{figures/selascfd.tex}}\\
\resizebox{0.5\textwidth}{!}{\input{figures/pelascfd.tex}}\\
\caption{CFD versus UFD phase shifts $\delta^I_\ell$ of $f^{1/2}_0(s)$ (top) and $f^{1/2}_1(s)$ (bottom).}
\label{fig:data}
\end{figure}

For our purposes, the most relevant partial wave is 
$f^{1/2}_0$ whose UFD is shown in Fig.~\ref{fig:data}. As explained in \cite{Pelaez:2016tgi} this wave is obtained by fitting 
 data measured in the $f^{1/2}_0+f^{3/2}_0/2$ and  $I=3/2$ combinations \cite{Cho:1970fb,Bakker:1970wg,Linglin:1973ci,Jongejans:1973pn,Estabrooks:1977xe,Aston:1987ir}. 
 It is also relevant that, as shown in Fig.\ref{fig:data}, 
 the $I=1/2$ vector wave UFD describes well the scattering data, in contrast to the solution \cite{DescotesGenon:2006uk}. The rest of the unconstrained partial-waves and high-energy input parameterizations are described in \cite{Pelaez:2016tgi} for $\pi K$ and \cite{Pelaez:2018qny} $\pi\pi\rightarrow K\bar K$. Minor updates will be detailed in a forthcoming publication \cite{inprep}.

\begin{figure}
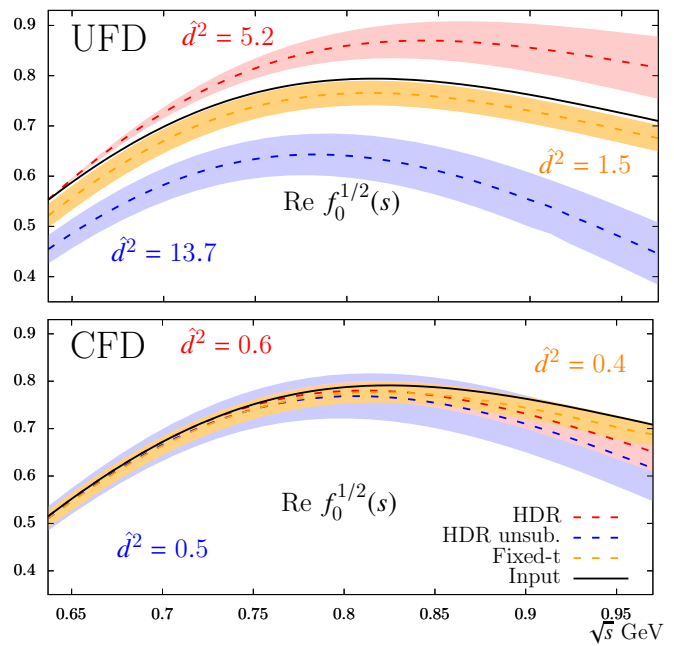

\resizebox{0.5\textwidth}{!}{\input{figures/kappa12ufd.tex}}
\resizebox{0.5\textwidth}{!}{\input{figures/kappa12cfd.tex}
}
\vspace*{0.1cm}
\caption{ Different dispersive outputs for the $f^{1/2}_0(s)$ partial wave, versus the input from the data parameterization.
Upper panel: Unconstrained Fits to Data (UFD). Note the huge discrepancies between the curves. Lower panel: Constrained Fits to Data (CFD). Now all curves agree within uncertainties. We only show the Roy Steiner results for $f^{1/2}_0(s)$ because these are the ones relevant for the $\kap$, but the CFD consistency is very good for the other dispersion relations and partial waves. Namely, the average $\chi^2/$dof per dispersion relation is 0.7, whereas the average $\chi^2/$dof is 1.4 per fitted partial wave.}
\label{fig:hdrufd}
\end{figure}

However, as seen in the upper panel of Fig.\ref{fig:hdrufd} when the UFD is used as input, the dispersive representation is not satisfied within uncertainties. Actually, the fixed-$t$ HDR output does not lie too far from the UFD input, but  the outputs using an unsubtracted or once-subtracted HDR for $F^-\equiv (F^{1/2}-F^{3/2})/3$ come on opposite sides and far 
from the input UFD parameterization. 

It is now very instructive to see how unstable is the pole parameter extraction from one fit that looks rather reasonable, as the UFD does. Thus, in Fig.\ref{fig:poles} we show the  \kap pole position calculated either using the HDR without subtractions for $F^-$ (hollow blue) or with one subtraction (hollow red). Note that {\it we use the same UFD input in the physical regions} but the two poles come out incompatible. This is mostly due to the pseudo-physical region of the $g^1_1$ partial wave.  The extraction would be even more unreliable if a simple model parameterization was used for the continuation to the complex plane 
instead of using a dispersion relation.

Thus, in order to obtain a rigorous and stable pole determination, we have imposed that the dispersive representation should be satisfied within uncertainties when fitting data. To this end, we have followed our usual procedure \cite{GarciaMartin:2011cn,Pelaez:2016tgi,Pelaez:2018qny}, defining a $\chi^2$-like distance 
$\hat{d}^2$ between the input and the output of each dispersion relation at many different energy values, which is then minimized together with the data $\chi^2$ when doing the fits. 

 We minimize simultaneously 16 dispersion relations. Two of them are the 
FDRs we already used in \cite{Pelaez:2016tgi}.
Four HDR are considered for the $\pi\pi \rightarrow K \bar{K}$
partial waves: Namely, once subtracted for $g^{0}_0$, $g^{0}_2,g^1_1$ and another unsubtracted for $g^1_1$, as we did in \cite{Pelaez:2018qny}. Note that here we also consider the once-subtracted case for $g^1_{1}$. 
In addition we now impose ten more dispersion relations within uncertainties for the $S$ and $P$ $\pi K$ partial-waves. 
Four of them come from fixed-$t$ and hyperbolic once-subtracted dispersion relations for $F^+\equiv(F^{1/2}+2F^{3/2})/3$, whereas the other six are two fixed-$t$ and another four HDR for $F^-$, either non-subtracted or once-subtracted. The HDRs applicability region in the real axis was maximized in \cite{Pelaez:2018qny}
choosing $a=-13.9 m_\pi^2$. We will however use here $a=-10 m_\pi^2$ as it still has a rather large applicability region in the real axis
and ensures that the \kap pole and its uncertainty fall inside the HDR domain. For the $\pi K$ $S^{1/2}$-wave, most of the dispersive uncertainty comes from the $\pi K$ $S$-waves themselves when using the subtracted $F^-$, whereas a large contribution comes from $g^1_1$ for the unsubtracted.

The details of our technique have been explained in \cite{Pelaez:2016tgi,Pelaez:2018qny}.
The resulting Constrained Fits to Data (CFD) differ slightly from the unconstrained ones, but still describe the data. This is illustrated in Fig. \ref{fig:data}, were we see that the difference between UFD and CFD is rather small for the $P$-wave, both providing remarkable descriptions of the scattering data. In contrast, in Fig. \ref{fig:data} we see that the CFD $S$-wave is lower than the UFD around and below the \kap nominal mass, but still describes well the experimental information. Also, Table~\ref{tab:lowpara} shows how the $S$-wave scattering lengths 
change from the UFD to the CFD. Note that our CFD values are consistent with previous dispersive predictions \cite{Buettiker:2003pp},  confirming some tension between data and dispersion theory versus recent lattice results \cite{Miao:2004gy,Beane:2006gj,Flynn:2007ki,Fu:2011wc,Sasaki:2013vxa,Helmes:2018nug}. Including those lattice values together with data leads to constrained fits satisfying dispersion relations substantially worse.

\begin{table}[!hbt] 
\caption{$S$-wave scattering lengths ($m_\pi$ units).}
\vspace{0.3cm}
\centering 
\begin{tabular}{l  c  c c} 
\hline
 & \hspace{0.2cm} UFD & \hspace{0.2cm} \bf{CFD} & \hspace{0.2cm} Ref. \cite{Buettiker:2003pp}\\
\hline\hline  
\rule[-0.2cm]{-0.1cm}{.55cm} $a^{1/2}_0$ & \hspace{0.2cm} 0.241$\pm$0.012 & \hspace{0.2cm} \bf{0.224$\pm$0.011} & \hspace{0.2cm} 0.224$\pm$0.022\\
\rule[-0.2cm]{-0.1cm}{.55cm} $a^{3/2}_0$ & \hspace{0.2cm} -0.067$\pm$0.012 & \hspace{0.2cm} \bf{-0.048$\pm$0.006} & \hspace{0.2cm} -0.0448$\pm$0.0077\\
\hline
\end{tabular} 
\label{tab:lowpara} 
\end{table}

Other waves suffer small changes from UFD to CDF, but are less relevant for the \kap (see \cite{inprep}). All in all, we illustrate in the lower panel of Fig.\ref{fig:hdrufd} that when the CFD is now used as input of the dispersion relations the curves of the input and the three outputs agree within uncertainties.

With all dispersion relations well satisfied we can now use our CFD as input in the HDR and look for the \kap pole. 
Results are shown in Fig.\ref{fig:poles}, this time as solid blue and red symbols depending on whether they are obtained with the unsubtracted or the once-subtracted $F^-$. Contrary to the UFD, the agreement between both determinations when using the CFD set is now remarkably good. Precise values of the pole position and residue for our subtracted and unsubtracted results 
are listed in Table~\ref{tab:poles}, together with
the dispersive result of \cite{DescotesGenon:2006uk} and our 
Pad\'e sequence determination \cite{Pelaez:2017sit}. 

Now, let us recall that the unsubtracted result depends strongly on the \pipikk pseudo-physical region, particularly on  $g^1_1$, whose unsubtracted dispersion relation error band is almost twice as big as the subtracted one. 
Moreover, under the change of $a$,
the subtracted result, both in the real axis and for the pole, barely changes, whereas the unsubtracted one only changes slightly (a few MeV for the pole). Therefore although both are compatible, we consider  the once-subtracted case more robust and thus our final result.

Let us remark that our dispersive pole obtained from data also agrees with the solution in \cite{DescotesGenon:2006uk}, although our width uncertainties are larger. In part, this is because we have estimated uncertainties for all our input.

For completeness, we have also calculated the parameters of the vector $K^*(892)$ pole, since we also describe the data there. We find: $\sqrt{s_p}=(890\pm 2) -i (25.6\pm 1.2)\,$MeV 
and its dimensionless residue $\vert g\vert= 5.69\pm0.12$. 

In summary, we have shown that simple, unconstrained fits to the existing $\pi K$ and $\pi \pi \rightarrow K \bar{K}$ data fail to satisfy hyperbolic and fixed-$t$ dispersion relations and yield rather unreliable \kap pole determinations. However, we have obtained  fits to data constrained to satisfy  those two hyperbolic dispersion relations, together with other Forward and fixed-$t$ dispersion relations. These constrained fits provide a rigorous, precise and robust determination of the \kap pole parameters. We think these results should provide the needed confirmation that, according to the Review of Particle Physics and the hadronic community, was needed to establish firmly the existence and properties of the \kap.

\begin{table}[!hbt] 
\caption{Poles and residues of the \kap. The last two lines are our dispersive outputs. The last line is our final result.}
\centering 
\begin{tabular}{l  c  c} 
\hline
\hline
 & \hspace{0.2cm} $\sqrt{s_{pole}}$\,\, (MeV) & \hspace{0.2cm} $|g|$\,\, (GeV) \\
\hline
\rule[-0.2cm]{-0.1cm}{.55cm} $K^*_0(700)$ \cite{DescotesGenon:2006uk}& \hspace{0.2cm} $(658\pm 13)-i (279\pm 12)$ & \hspace{0.2cm} ---\\
\rule[-0.2cm]{-0.1cm}{.55cm} $K^*_0(700)$ \cite{Pelaez:2016klv}& \hspace{0.2cm} $(670\pm 18)-i (295\pm 28)$ & \hspace{0.2cm} 4.4$\pm$0.4\\
\rule[-0.2cm]{-0.1cm}{.55cm} $K^*_0(700)$ 0-sub & \hspace{0.2cm} $(648\pm 6)-i (283\pm 26)$ & \hspace{0.2cm} 3.80$\pm$0.17\\
\rule[-0.2cm]{-0.1cm}{.55cm} {\bf $\bf{K^*_0(700)}$ 1-sub}& \hspace{0.2cm}  $\bf{(648\pm 7) -i (280\pm 16)}$ & \hspace{0.2cm} $\bf{3.81 \pm 0.09}$\\
\hline
\hline
\end{tabular} 
\label{tab:poles} 
\end{table}

{\bf Acknowledgments} 

 This project has received funding from the Spanish MINECO grant FPA2016-75654-C2-2-P and the European Union’s Horizon 2020 research and innovation programme under grant agreement No 824093 (STRONG2020). AR would like to acknowledge the financial support of the U.S. Department of Energy contract DE-SC0018416 and of the Universidad Complutense de Madrid.

\bibliographystyle{apsrev4-1}
\bibliography{biblio.bib}

\end{document}